# Designing order-disorder transformation in high-entropy ferritic steels

Prashant Singh[a] and Duane D. Johnson [a,b]

[a] Ames Laboratory, United States Department of Energy, Ames, Iowa 50011, USA
[b] Department of Materials Science & Engineering, Iowa State University, Ames, Iowa 50011, USA

**Abstract**

Order-disorder transformations hold an essential place in chemically complex high-entropy ferritic-steels (HEFSs) due to their critical technological application. The chemical inhomogeneity arising from mixing of multi-principal elements of varying chemistry can drive property altering changes at the atomic scale, in particular short-range order. Using density-functional theory based linear-response theory, we predict the effect of compositional tuning on the order-disorder transformation in ferritic steels – focusing on Cr-Ni-Al-Ti-Fe HEFSs. We show that Ti content in Cr-Ni-Al-Ti-Fe solid solutions can be tuned to modify short-range order that changes the order-disorder path from BCC-B2 (Ti atomic-fraction = 0) to BCC-B2-L2$_1$ (Ti atomic-faction > 0) consistent with existing experiments. Our study suggests that tuning degree of SRO through compositional variation can be used as an effective means to optimize phase selection in technologically useful alloys.

**Introduction**

High-entropy alloys, including metals and ceramics with near-equiatomic compositions with four and more elements [**1-6**], continue to gain significant interest due to the unprecedented opportunity to explore large materials design space and uncover potentially remarkable compositions with outstanding structural and functional properties [**4-11**]. The design strategy in high-entropy alloys has been to use the concept of entropy to stabilize the single-phase solid-solution (e.g., face-centered cubic (FCC phase) or body-centered cubic (BCC phase)) [**12**] with an attempt to find specific electronic, thermodynamic, and microstructural properties [**4,13-15**]. While the progress over the last decade towards first-generation high-entropy alloys is remarkable, the critical thermodynamic behavior of these alloys indicates that only a little is known on the effects of SRO [**16**] and the associated lattice deformations [**17**] on electronic and/or mechanical response [**18**].

High-entropy ferritic steels (HEFSs) are one important class of multi-principal element alloys due to their cost efficiency, low thermal expansion, and good thermal conductivity compared to Ni-based superalloys and austenitic steels [**19-21**]. Similar to conventional alloys, HEFSs show chemistry and temperature dependent ordering that may undergo one or more phase transitions into less ordered phases. Precipitation hardening due to presence of ordered phases in HEFSs gives excellent creep and oxidation behavior [**22**], which is analogous to the presence of $\gamma'$ phase in Ni-based superalloys. Unlike $L1_2$ phases in austenitic steels, presence of ordered B2 or $B2/L2_1$ phases in body-centered cubic matrix [**23,24**] may provide similar mechanical effects [**25**]. Therefore, a detailed understanding of order-disorder transformations and precipitate formation along with compositional control in HEFSs can be of fundamental importance.

Here we present a systematic study on the effect of compositional tuning of Ti on order-disorder behavior in Cr-Ni-Al-Ti-Fe HEFSs using density-functional theory (DFT) methods in combination with configurational averaging [**26,27**]. The linear response theory was used for calculating short-range order in the disordered Cr-Ni-Al-Ti-Fe HEFSs [**25**]. We show that degree of SRO can be controlled using Ti content, which modifies order-disorder pathway from BCC-B2 to BCC-B2-L2$_1$ [**28**]. Our findings are in good agreement with recent observations of Wolf-Goodrich *et al.* [**29**], who report combinations of BCC/B2 and BCC/B2/L2$_1$ phases depending on Ti composition (at.%Ti) in Cr-Ni-Al-Ti-Fe HEFSs. The linear-response theory for SRO analyzed by concentration wave method was used to seed the fully self-consistent KKR-CPA calculation in the broken symmetry case, which, unlike Monte-Carlo methods [**30**], does not rely on fitted interactions. We also discussed the phase stability (formation energy) and electronic-structure origin of disordered and (partial) ordered phases for selected HEFS compositions. We found that SRO can be a key structural feature for optimizing phase selection and mechanical response.

**Computational details:**

**Density-functional based linear-response theory:** Phase stability and electronic-structure were addressed using an all-electron, Green's function based Korringa-Kohn-Rostoker (KKR) electronic-structure method [**26**]. The configurational averaging to tackle chemical disorder is handled using the coherent-potential approximation (CPA) [**27**], and the screened-CPA was used to address Friedel-type charge screening [**31**]. Valence electrons and shallow lying core electrons affected by alloying are addressed via a scalar-relativistic approximation (where spin-orbit terms only are ignored) [**26,27,31**], whereas deep lying core are address using the full Dirac solutions. Electronic density of states (DOS) and Bloch-spectral function (BSF) were calculated within the atomic sphere approximation (ASA) with periodic boundary conditions. The interstitial electron contributions to Coulomb energy are incorporated using Voronoi polyhedra. The generalized gradient approximation to DFT exchange-correlation was included using the *libXC* opensource code [**32**]. Brillouin-zone integrations for self-consistent charge iterations were performed using a Monkhorst-Pack *k*-point mesh [**33**]. Each BSF was calculated for 300 k-points along high-symmetry lines in an irreducible Brillouin zone.

**Thermodynamic linear-response theory for short-range order:** Chemical short-range order and associated instabilities were calculated using KKR-CPA-based thermodynamic linear-response theory [**28,34-37**]. The Warren-Cowley SRO (pair-correlation) parameters $\alpha_{\mu\nu}(\mathbf{k};T)$ for µ-v elemental pairs are calculated directly in Laue units [**28**]. The necessary energy integrals over the Green's functions were performed at finite temperature by summing over Matsubara frequencies [$\omega_n$ = k$_B$T(2$n$+1)$\pi$] [**28**]. Dominant pairs driving SRO are identified from the chemical pair-interchange energies $S^{(2)}_{\mu\nu}(\mathbf{k};T)$ (a thermodynamically averaged quantity – not a pair interaction), determined from an analytic second-variation of the DFT-based KKR-CPA grand potential with respect to concentrations fluctuations of $c^i_\mu$ at atomic site *i* and $c^j_\nu$ at atomic site *j* [**28**]. The chemical stability matrix $S^{(2)}_{\mu\nu}(\mathbf{k}; T)$ reveals the unstable Fourier modes with ordering wavevector k$_o$, or clustering if at k$_o$=(000) at spinodal temperature (T$_{sp}$) [**28**]. Here, T$_{sp}$ is the temperature where SRO diverges, i.e., $\boldsymbol{\alpha}^{-1}_{\mu\nu}(\mathbf{k}_o;T_{sp})$=0, which signifies absolute instability in alloy and provides an estimate of order-disorder (ordering systems) or miscibility gap (in clustering systems).

**Formation energy calculation**: Formation energy (E$_{form}$) of the Cr-Ni-Al-Ti-Fe HEFSs was estimated using $\mathrm{E_{form} = E_{total}^{Cr-Ni-Al-Ti-Fe}(c_i) - \sum_{i=1,N} c_i E_i}$, where $\mathrm{E_{total}^{Cr-Ni-Al-Ti-Fe}}$ is the total energy, $\mathrm{c_i}$ is elemental composition, $\mathrm{E_i}$ is the energy of alloying elements, and '*i*' labels elements BCC (Cr, Fe) FCC (Ni, Al), HCP Ti.

**Temperature Estimates**: The Curie (or ferromagnetic ordering) temperature of Cr-Ni-Al-Ti-Fe HEFSs was assessed using mean-field Heisenberg-like model [**38**]. The mean-field relation for Tc $\left[=\frac{2}{3}[E_{DLM} - E_{FM}]/k_B\right]$ is proportional to the energy difference between paramagnetic (PM) and ferromagnetic (FM) states, with the PM state approximated by the disordered local moment (DLM) state – randomly oriented (uncorrelated) local moments easily represented with a separate CPA condition for moment orientations [**39**]. However, as discussed by Sato *et al.* [**38**] for dilute magnetic semiconductor, it is appropriate to consider a slightly modified relation as an upper bound, i.e., $T_c = \frac{2}{3} \cdot \left[\frac{1}{1-c}\right] \cdot [E_{DLM} - E_{FM}]/k_B$, given the Cr-Ni-Al-Ti-Fe HEFSs has a non-magnetic element (Al) with concentration *c*.

**Concentration (Fourier) Wave analysis**: Fourier analysis or concentration wave approach was used to interpret the (partial)long-range order observed in the SRO calculations, where normal modes ($e_i^\sigma$) in Gibbs' space were obtained from chemical stability matrix in linear-response theory [**28**]. The occupation probabilities [$n_i(\mathbf{r})$] at site $\mathbf{r}_i$ are identical to elemental compositions [$c_i$] in disorder phase of the alloy, which depends on type of order in long-range ordered phase. Here, 'i' is the index for the type of elements. The occupation probabilities in N-component system, i.e., HEFSs, can be expanded in as Fourier series, i.e., concentration wave, which can be written in terms of normal modes as

$$\begin{bmatrix} n_1(\mathbf{r}) \\ n_2(\mathbf{r}) \\ n_3(\mathbf{r}) \\ \ldots \\ n_{N-1}(\mathbf{r}) \end{bmatrix} = \begin{bmatrix} c_1 \\ c_2 \\ c_3 \\ \ldots \\ c_{N-1} \end{bmatrix} + \sum_{s,\sigma} \eta_\sigma^s \begin{bmatrix} e_1^\sigma(\mathbf{k}_s) \\ e_2^\sigma(\mathbf{k}_s) \\ e_3^\sigma(\mathbf{k}_s) \\ \ldots \\ e_{N-1}^\sigma(\mathbf{k}_s) \end{bmatrix} \times \sum_{j_s} \gamma^\sigma(\mathbf{k}_{j_s}) e^{i\mathbf{k}_{j_s} \cdot \mathbf{r}} \quad \text{Eq. (1)}$$

For a given atomic position $r_i$, $c_i$ is the composition vector of order (N–1) component, relative to "host" element N. The sum in Eq. (1) runs over the star of inequivalent wavevectors 's' that defines the order, $\sigma$ is eigenvector branch of the free-energy quadric, and $j_s$ are equivalent wavevectors in star *s*. The $\eta_\sigma^s$ ($0 \leq \eta(T) \leq 1$) is long-range order parameter of star `s' and branch `$\sigma$; where $e_i^\sigma(\mathbf{k})$ is eigenvector of the normal concentration mode for branch $\sigma$, and symmetry coefficients $\gamma^\sigma(\mathbf{k}_{j_s})$ found by normalization condition and lattice geometry. Note that the elements of vectors $n_i(\mathbf{r})$ and $c_i$ conserve probability and must add to 1 ($\sum_{i=1}^{N} c_i = 1$), i.e., by the sum rule first (N – 1) elements should add to final elements.

**Results and Discussion**

Phase stability, structural, and magnetic property analysis for Cr-Ni-Al-Ti-Fe HEFSs, related to recent experimental work of Wolf-Goodrich *et al.* [**29**], are shown in **Table 1**. We calculated the formation energy (**E$_{form}$**) of each HEFSs in BCC, FCC and HCP phases. The calculated formation energy in HCP phase for each alloy was a large positive number compared to BCC and FCC phases, therefore, not discussed. Our phase stability analysis **Table. 1** indicates that the BCC phase is energetically more favorable in Cr-Ni-Al-Ti-Fe HEFSs. This is the reason, we mainly focused on BCC Cr-Ni-Al-Ti-Fe HEFSs. The DFT calculated formation energies in **Table 1** show increase in E$_{form}$ with increase in Ti+Al composition, where Cr$_{0.05}$Ni$_{0.15}$Al$_{0.30}$Ti$_{0.15}$Fe$_{0.35}$ HEFS with Ti+Al=0.45 at.-frac. was found energetically more favorable compared to other alloys.

The trends in magnetization (cell moment) and Curie-temperature (estimated using mean-field Heisenberg-like model [**38**]) show increase with decreasing Ti+Al+Cr composition in **Table 1** (Cr composition was included with Ti+Al in our magnetization analysis as anti-ferromagnetic character of Cr is well known to impact the magnetic behavior of the alloy). The total moment was found to decrease with increasing Ti+Al+Cr composition from 0.56 $\mu_B$ (Ti+Al+Cr=0.50 at.-frac.) to 0.10$\mu_B$ (Ti+Al+Cr=0.60 at.-frac.). To understand this better, a detailed local moment analysis was performed on $Cr_{0.20}Ni_{0.10}Al_{0.30}Fe_{0.40}$ (no Ti) and $Cr_{0.20}Ni_{0.10}Al_{0.25}Ti_{0.10}Fe_{0.35}$ (with Ti) HEFSs. The local moment at Cr was found to increase from -0.25 $\mu_B$ (no Ti) to -0.41 $\mu_B$ (with Ti), respectively. The sign of local moment shows that Cr prefers to align anti-ferromagnetically (AFM) irrespective of the initial orientation (FM or AFM) compared to other magnetic species such as Ni and Fe. The frustrated moment at Cr (AFM arrangement) subsidizes the overall cell moment in HEFSs with increasing Cr composition. We found that higher Ti composition weakens the AFM (frustration) nature of Cr, which reduces the overall magnetic strength.

Regarding structural property, for example, bulk moduli (K), no significant change was observed with Ti+Al composition variation. We observed in our calculations that Al+Ti or Ti stabilizes the BCC phase over FCC phase, which is obvious from the calculated formation energies in **Table. 1**. For example, equiatomic quaternary $Cr_{0.25}Ni_{0.25}Al_{0.25}Fe_{0.25}$ HEFSs (with no Ti) shows positive formation enthalpy, whereas alloys in presence of Ti show improved BCC phase stability.

**Table 1.** Lattice-constant (Å), formation energy (meV/atom), bulk-moduli (GPa), magnetization ($\mu_B$/cell), and Curie temperature (K) for various Cr-Ni-Al-Ti-Fe HEFSs.

| HEFSs | a [Å] | $E_{form}$ [meV/atom] | | | $K_{BCC}$ [GPa] | Mag. [$\mu_B$/cell] | $T_c$ [K] |
|---|---|---|---|---|---|---|---|
| | | BCC | | FCC | | | |
| | | FM | PM | FM | | | |
| CrNiAlTiFe | 2.92 | -0.41 | -0.27 | 80.01 | 162.5 | 0.10 | 1.8 |
| $Cr_{0.20}Ni_{0.10}Al_{0.25}Ti_{0.10}Fe_{0.35}$ | 2.89 | -12.27 | 8.78 | 83.21 | 173.8 | 0.38 | 250.5 |
| $Cr_{0.05}Ni_{0.15}Al_{0.30}Ti_{0.15}Fe_{0.35}$ | 2.93 | -82.79 | -59.26 | 19.28 | 173.8 | 0.43 | 330.9 |
| $Cr_{0.20}Ni_{0.10}Al_{0.30}Fe_{0.40}$ | 2.87 | -21.42 | 10.13 | 58.14 | 169.7 | 0.56 | 348.7 |
| CrNiAlFe | 2.86 | 7.82 | 19.92 | 58.75 | 174.8 | 0.33 | 124.9 |

The thermodynamic stability of multicomponent alloys is an important criterion to understand relative phase stability with respect to alloying element, which requires non-trivial sampling over infinitely large configurations in disorder phase [26,27]. Recently, Singh *et al* [34,37] extended the Hume-Rothery criteria [36,40] by including DFT (KKR-CPA) calculated formation energies (including proper configuration averaging [26,27]) along with (i) size-effect, (ii) lattice structure, (iii) valence-electron composition (VEC), and (iv) electronegativity difference. Instead of empirically estimated formation-energies, the inclusion of $E_{form}$ ($DFT$) has made the design criteria of predicting phase stability more robust. To better understand the effect of alloying elements on thermodynamic stability and structural behavior of Cr-Ni-Al-Ti-Fe HEFSs, we calculated and present $E_{form}$ (in **Fig. 1**) and (bulk-moduli (K), volume (V), lattice constant (a) in **Fig. 2a-c**) with respect to each alloying elements as a line plot, i.e., $Cr_x(NiAlTiFe)_{1-x}$, $Ni_x(CrAlTiFe)_{1-x}$, $Al_x(NiCrTiFe)_{1-x}$, $Ti_x(NiAlCrFe)_{1-x}$, and $Fe_x(NiAlTiCr)_{1-x}$.

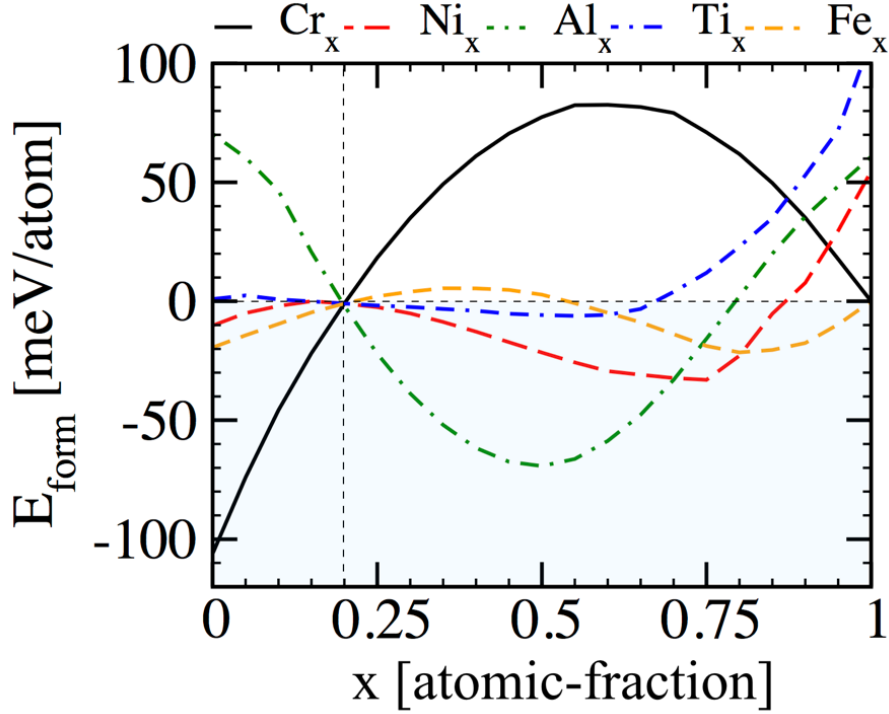

**Figure 1.** The formation energies ($E_{form}$; in mRy/atom) of $Cr_x(NiAlTiFe)_{1-x}$, $Ni_x(CrAlTiFe)_{1-x}$, $Al_x(NiCrTiFe)_{1-x}$, $Ti_x(NiAlCrFe)_{1-x}$, and $Fe_x(NiAlTiCr)_{1-x}$ HEFSs. The equiatomic high-entropy composition is shown by vertical dashed line.

In **Fig. 1**, we plot $E_{form}$ in Cr-Ni-Al-Ti-Fe HEFSs, where shaded zone below horizonal line shows the energetically stable alloy compositions, while the vertical dashed line is the equiatomic high-entropy composition. The solid black line shows the variation of $E_{form}$ with respect to Cr composition. The solubility limit of Cr in Ni-Al-Ti-Fe HEFSs is 0-0.20 atomic-fraction (at.-frac.) beyond which the alloy becomes energetically unstable. On the other hand, it was found that the solubility limit of Ni in Cr-Ni-Al-Ti-Fe with two different zones, (i) 0-0.15 at.-frac., and (ii) 0.32-0.85 at.-fac., i.e., Ni in composition range 0.15-0.32 at.-frac. remains weakly stable or unstable for forming BCC HEFSs. The solubility limit (energy stability) of Al was found from 0.25-to-0.80 at.-frac., which shows that stability of BCC $Al_x(NiCrTiFe)_{1-x}$ increases with increasing Al at.-frac. We also found that, although Ti solubility range is from 0-0.65 at.- frac. has very weak effect on energy stability, i.e., no major benefits of adding excess Ti in alloy. Interestingly, Fe is more peculiar here, because varying Fe at.-frac. in Cr-Ni-Al-Ti-Fe HEFSs shows strong solubility of Fe from 0.60-1 at.f-rac. Based on our stability analysis, we found that optimal composition ranges for elements like Cr, Ni, Al, and Ti are (0-0.20), (0-0.15), (0.25-0.80), and (0-0.20) at.-frac., respectively. The optimal range found in our phase stability analysis clearly matches with compositions experimentally synthesized by Wolf-Goodrich *et al.* [**29**], where author's report the formation of different type of precipitates in BCC alloys such as B2/L2$_1$ depending on specific elemental compositions.

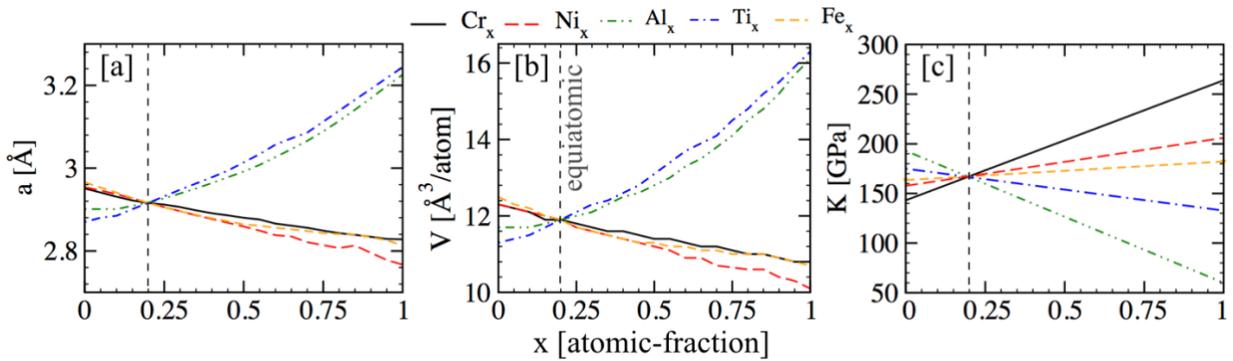

**Figure 2.** (a) Lattice constant (Å), (c) volume (V, in Å$^3$-atom$^{-1}$), and (c) bulk-moduli (K, in GPa) of Cr$_x$(NiAlTiFe)$_{1-x}$, Ni$_x$(CrAlTiFe)$_{1-x}$, Al$_x$(NiCrTiFe)$_{1-x}$, Ti$_x$(NiAlCrFe)$_{1-x}$, and Fe$_x$(NiAlTiCr)$_{1-x}$ HEFSs. The K in (c) was fitted with volume and total-energy using Birch-Murnaghan equation of state.

For small strains, the relation between lattice constant (a) and volume (V) with bulk moduli (K) can be defined in terms of change in lattice constant with respect to a ($\Delta a/a$), change in pressure ($\Delta P$) required for volume change per unit volume, i.e., $\frac{\Delta a}{a} = \frac{\Delta P}{3K}$. The expression suggests that change in lattice constant inversely related to change in volume or lattice constant as shown in **Fig. 2a-c**. We can see in Fig. **2a-b** that `a' and `V' increases with increase in Al/Ti compositions whereas decreases for Cr/Ni/Fe.

Clearly, bulk moduli in **Fig. 2c** clearly decreases for Al/Ti whereas increases for Cr/Ni/Fe cases that agrees well with relationship between a/V with K in $\frac{\Delta a}{a} = \frac{\Delta P}{3K}$.

To elucidate chemical ordering in Cr-Ni-Al-Ti-Fe HEFSs, we calculated SRO on Ti-rich and Ti-poor compositions, i.e., $Cr_{0.20}Ni_{0.10}Al_{0.25}Ti_{0.10}Fe_{0.35}$ and $Cr_{0.20}Ni_{0.10}Al_{0.30}Fe_{0.40}$ (no Ti). While KKR-CPA **E**$_{form}$ determines ground-state stability of BCC versus FCC, our linear-response SRO calculations indicate [**28**] directly the chemical instabilities, i.e., clustering or ordering modes, inherent in a given high-entropy alloys [**28,34,37**], and the likely low-temperature long-range order [**28**], including its electronic origin.

In **Fig. 3a-d**, we show SRO and $S^{(2)}_{\alpha\beta}$(**k**, T) for $Cr_{0.20}Ni_{0.10}Al_{0.25}Ti_{0.10}Fe_{0.35}$ and $Cr_{0.05}Ni_{0.15}Al_{0.30}Ti_{0.15}Fe_{0.35}$ HEFSs, respectively. **Figure 3a,c** reveals a decrease in local chemical order to from 6 Laue for $Cr_{0.20}Ni_{0.10}Al_{0.25}Ti_{0.10}Fe_{0.35}$ (**Fig. 3a**) to 3 Laue for $Cr_5Ni_{15}Al_{30}Ti_{15}Fe_{35}$ (**Fig. 3c**) with decrease in Ti (0.15 to 0.10 at.-frac.) composition. The chemical stability matrix [$S^{(2)}_{\alpha\beta}$(**k**, T=1.15T$_{sp}$)] plot in **Fig. 3b,d** reveals the atomic pairs and modes driving SRO that are manifested in SRO pairs in **Fig. 3a,b**. The SRO at wavevector **k=k$_o$** reveals maximal diffuse intensities above the spinodal temperature T>T$_{sp}$, i.e., 895 K for 0.10 Ti at.-frac. in $Cr_{0.20}Ni_{0.10}Al_{0.25}Ti_{0.10}Fe_{0.35}$ and 815 K for 0.15 Ti at.-frac. in $Cr_{0.05}Ni_{0.15}Al_{0.30}Ti_{0.15}Fe_{0.35}$. The SRO in BCC phase for both the alloys in **Fig. 3a,b** shows maximal peak at H=(111) point that indicates B2 type ordering with a potentially secondary ordering mode at P=(½ ½ ½). The presence of combined H+P type ordering peaks indicate L2$_1$-type ordering [**37**]. Below T$_{sp}$, the SRO predicts possible phase decomposition into B2 of disordered $Cr_{0.20}Ni_{0.10}Al_{0.25}Ti_{0.10}Fe_{0.35}$ and $Cr_{0.05}Ni_{0.15}Al_{0.30}Ti_{0.15}Fe_{0.35}$ HEFSs, and the B2 phase may undergo secondary phase transformation into L2$_1$ on further lowering the temperature. The atomic pairs in $S^{(2)}(\mathbf{k}, 1.15T_{sp})$ in **Fig. 3b,d** shows that the Al-Fe and Al-Ti pairs are the strongest pairs driving phase decomposition below spinodal temperature, i.e., T$_{sp}$ of 895 K for Cr$_{0.20}$Ni$_{0.10}$Al$_{0.25}$Ti$_{0.10}$Fe$_{0.35}$ and 810 K for Cr$_{0.05}$Ni$_{0.15}$Al$_{0.30}$Ti$_{0.15}$Fe$_{0.35}$, respectively. We note that driving modes in $S^{(2)}(\mathbf{k}, 1.15T_{sp})$ at H- and P-points are different, however, the dominant SRO modes at (H+P)-point in both **Fig. 3a** and **Fig. 3c** are the same, i.e., Ni-Al and Al-Ti pairs. This asymmetry atomic pairs in SRO and $S^{(2)}$ pairs occur due to the conservation of the sum rule [**28**].

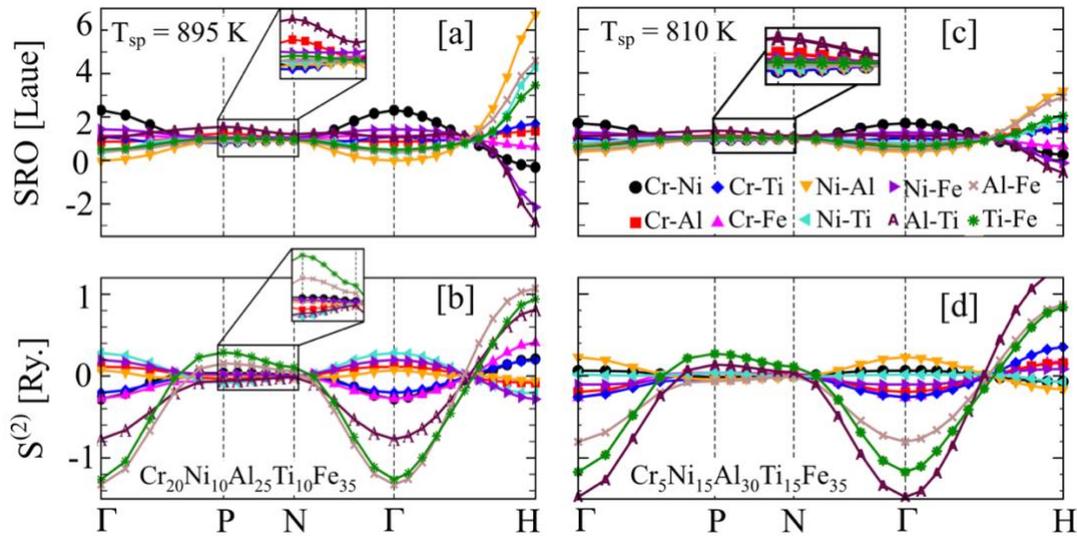

**Figure 3**. (a,c) SRO (in Laue), and (b,d) chemical stability matrix $S^{(2)}_{\alpha\beta}(\mathbf{k}, T=1.15T_{sp})$ [in Rydberg] for Cr$_{0.20}$Ni$_{0.10}$Al$_{0.25}$Ti$_{0.10}$Fe$_{0.35}$ and Cr$_{0.05}$Ni$_{0.15}$Al$_{0.30}$Ti$_{0.15}$Fe$_{0.35}$ HEFSs, respectively, along high-symmetry directions (Γ-P-N-Γ-H) of BCC Brillouin zone. Peaks at H=(100) indicate B2-type SRO (dominated by **Al-Ni pairs**). Secondary peaks at Γ and P suggest possible segregation (dominated by **Cr-Ni pair**) and weaker L2$_1$-type SRO (dominated by **Al-Ti pair**).

The Block-spectral function (BSF) and partial density of states (PDOS) for BCC Cr$_{0.20}$Ni$_{0.10}$Al$_{0.25}$Ti$_{0.10}$Fe$_{0.35}$ are shown in **Fig. 4a,b**. The BSF in **Fig. 4a** shows large disorder broadening (at)near Fermi energy (E$_{Fermi}$), where scale on right shows weak (black) to strong (red) disorder effect arising from mixing of different alloying species [33]. For in-depth understanding of the alloying effect, we plot the PDOS of $Cr_{0.20}Ni_{0.10}Al_{0.25}Ti_{0.10}Fe_{0.35}$ in **Fig. 4b**. The four distinct energy regions are shaded at (i) -0.052 mRy, (ii) -0.094 mRy, (iv) -0.20 mRy, and (iv) -0.275 mRy below $E_{Fermi}$ in PDOS. (i) The energy region shaded in pink shows strong hybridization among overlapping peaks of Cr-*3d*, Fe-*3d*, and Ti-*3d* states, which also coincides with strongly diffused BSF along $\Gamma$-H in **Fig. 4a**. Similarly, regions (ii-iv) in **Fig. 4b** show strong hybridization among (ii) Cr-*3d*, Fe-*3d*, Al-*2p* in blue region, (iii), Ni-*3d*, Al-*2p* in orange region, and (iv) Cr-*3d*, Fe-*3d*, Ni-*3d*, Al-*2p* in green region), respectively. We some obvious instances in regions ii-iv, where the transition-metal *d*-states and the Al-p states were found to show strong hybridization. The strongly diffused bands (in red) in BSF in the energy range (ii-iv) also agrees well with PDOS analysis. The strong hybridization found between Ni-*3d* and Al-*2p* bands at -0.20 mRy in **Fig. 4b** directly connects with dominant SRO pair with H-point ordering in **Fig. 3**.

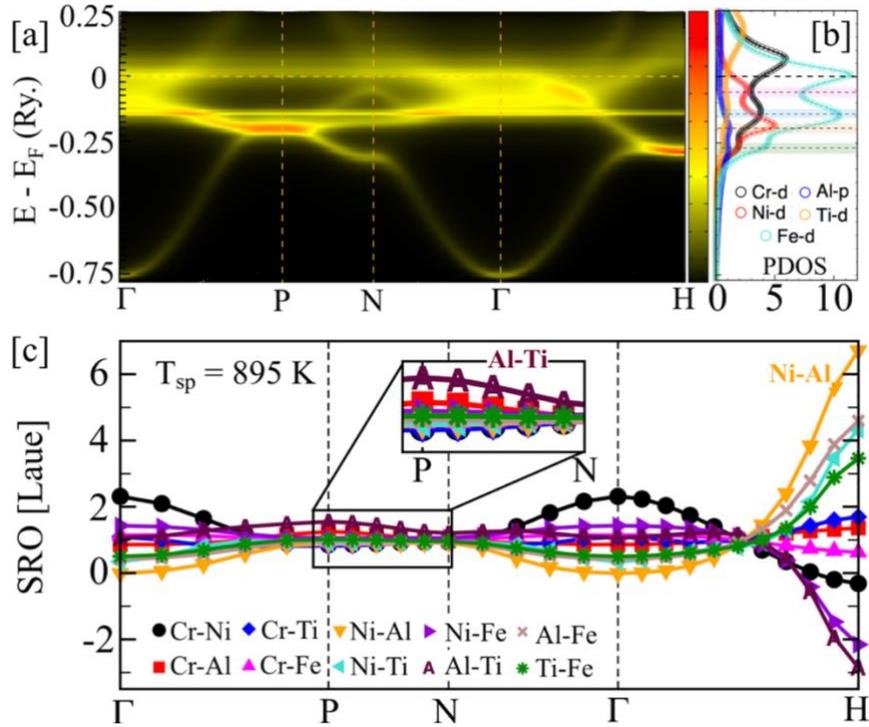

**Figure 4**. (a) Bloch-spectral functions (i.e., electronic dispersion, with broadening due to chemical disorder), (b) partial density of states, and (c) short-range order (in Laue) along high-symmetry directions (Γ-P-N-Γ-H) of a BCC Brillouin zone. Peaks at H=(111) indicate B2-type SRO dominated by Al-Ni pair. The possibility of secondary ordering phase arises due to presence of stronger ordering peak at H in Ni-Al pair and weaker ordering peak at P in Al-Ti pair. The SRO peaks at H+P are suggestive of L2$_1$ phase [**25**].

Indeed, the expectation of low-temperature ordering due to increased hybridization among alloying elements was also confirmed by the presence of strong SRO peaks at H-point (indicating B2-type mode) and H+P-point (indicating L2$_1$-type mode) for Cr$_{0.20}$Ni$_{0.10}$Al$_{0.25}$Ti$_{0.10}$Fe$_{0.35}$ HEFS in **Fig. 4c**. The maximal SRO peak at H-point in **Fig. 4c** shows Ni-Al dominated B2-type ordering. The strong Ni-Al peak at H-point is followed by Fe-Al and Ni-Ti SRO pairs. A fairly strong Fe-Al SRO can be attributed to the larger solubility of Fe than Cr at low-temperature in ordering phases. Also, a weaker Cr-Ni peak in SRO at Γ-point in **Fig. 4c** indicates the tendency of segregation, i.e., energetically Cr and Ni do not prefer same neighboring environment [**28**]. Note that a weak secondary Al-Ti peak at P-point as shown in inset of **Fig. 4c** is indicative of Ti enriched B2 phase. The presence of a weak ordering peak at P along with strong ordering peak at P is consistent with coexistant B2 and L2$_1$ phases as reported by Wolf-Goodrich *et al.* [**29**] in nearly same composition as Cr$_{0.20}$Ni$_{0.10}$Al$_{0.25}$Ti$_{0.10}$Fe$_{0.35}$ HEFS.

To explore the possibility of coexistent phases, we extracted H- and P-point eigenvectors from the SRO analysis above phase decomposition temperature for analytically solving CW Eq. (1) [**37**]. The CW analysis shows that partially ordered B2 and L2$_1$ phases of Cr$_{0.20}$Ni$_{0.10}$Al$_{0.25}$Ti$_{0.10}$Fe$_{0.35}$ HEFS exhibit lower energies than BCC by −35.22 meV-atom$^{-1}$ and −53.72 meV-atom$^{-1}$, respectively. Notably, Amalraj *et al.* [**41**] also observed L2$_1$ peaks in Cr$_{0.20}$Ni$_{0.10}$Al$_{0.25}$Ti$_{0.10}$Fe$_{0.35}$. Hence, the phases (and their estimated energy gains) initially indicated by the SRO in $Cr_{0.20}Ni_{0.10}Al_{0.25}Ti_{0.10}Fe_{0.35}$ in **Fig. 3a** (same in **Fig. 4c**) shows a good agreement with recent experiments [**29,41**].

The SRO in Ti-rich ($Cr_{0.20}Ni_{0.10}Al_{0.25}Ti_{0.10}Fe_{0.35}$) and no-Ti ($Cr_{0.20}Ni_{0.10}Al_{0.30}Fe_{0.40}$) cases in **Fig. 5a-b** were compared to better understand the effect of Ti. We found that pronounced secondary-ordering peak at P-point (Al-Ti pair; see inset) in **Fig. 5a** disappears when Ti is reduced to 0 at.-frac. in $Cr_{0.20}Ni_{0.10}Al_{0.30}Fe_{0.40}$ HEFS in **Fig. 5b**. Unlike $Cr_{0.20}Ni_{0.10}Al_{0.25}Ti_{0.10}Fe_{0.35}$ (with strong B2 (H-point) ordering along with possible $L2_1$ ordering in **Fig. 5a**), the $Cr_{0.20}Ni_{0.10}Al_{0.30}Fe_{0.40}$ HEFS only shows possible B2 ordering with no-sign of $L2_1$ (in strong agreement with observations of Wolf-Goodrich *et al.* [**29**]).

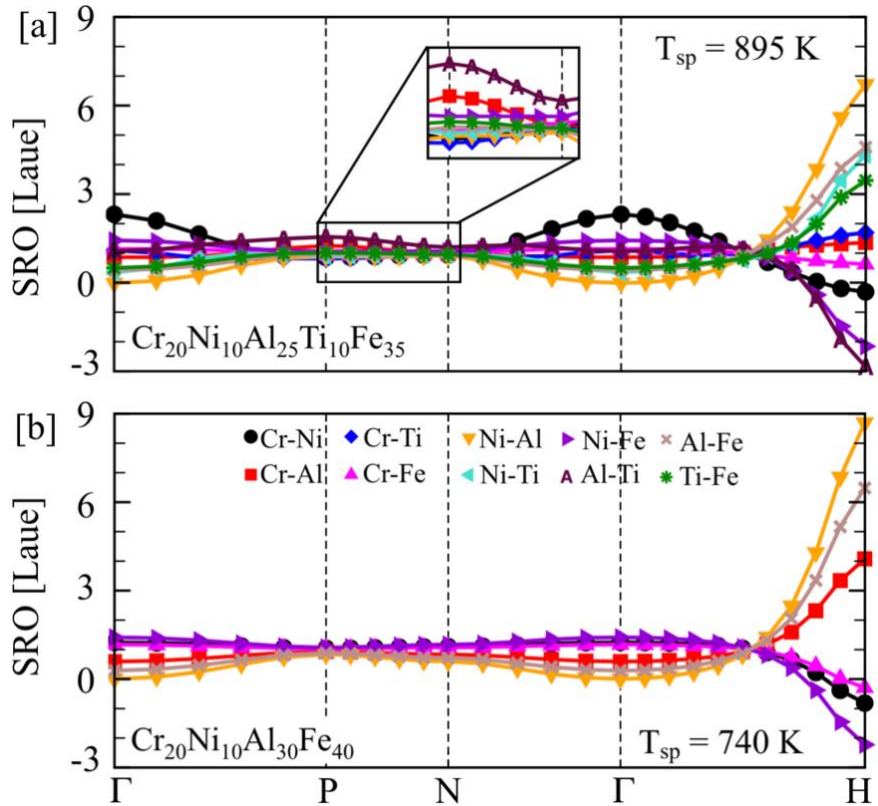

**Figure 5**. The short-range ordering pairs for (a) $Cr_{0.20}Ni_{0.10}Al_{0.25}Ti_{0.10}Fe_{0.35}$ and (b) $Cr_{0.20}Ni_{0.10}Al_{0.30}Fe_{0.40}$ HEFSs are shown along high-symmetry directions ($\Gamma$-P-N-$\Gamma$-H) of BCC Brillouin zone. The potential secondary ordering peak at P ($L2_1$) in (b) disappears at Ti=0 atomic-fraction.

The direct energy calculation of low temperature ordering phases (B2 and L2$_1$) and comparing them with disorder phase (BCC) will allow us to establish the fact that predicted incipient long-range order phases may exist [**28**]. However, the determination sublattice occupation (on phase decomposition of disorder phase) is needed for energy calculation of ordering phases using DFT, which remains unknown. The concentration wave in Eq. (1) [**42,44**] was used to estimate the occupation probabilities of each alloying element in possible ordering phases (B2 and L2$_1$ are two such possibilities in disorder BCC phase) of Cr-Ni-Al-Ti-Fe HEFSs, which is required by DFT for direct energy calculation.

The CW Eq. (1) was rewritten using the eigen-vector information extracted from SRO analysis of Cr$_{0.20}$Ni$_{0.10}$Al$_{0.25}$Ti$_{0.10}$Fe$_{0.35}$ HEFS below the phase decomposition temperature as

$$\begin{bmatrix} n(Cr) \\ n(Ni) \\ n(Al) \\ n(Ti) \end{bmatrix} = \begin{bmatrix} 0.20 \\ 0.10 \\ 0.25 \\ 0.10 \end{bmatrix} + \frac{\eta}{2} \begin{bmatrix} 0.3550 \\ 0.3955 \\ 0.7603 \\ -0.3736 \end{bmatrix} \exp[2 \times \pi \times i \times \boldsymbol{r} \times (111)] \quad \text{Eq. (2)}$$

here, the CW equation is solved for unknown order parameter `$\eta$'. The factor $\gamma = \frac{1}{2}$ comes from the symmetry of the BCC cell with two lattice cites, i.e., r= (000) and (½ ½ ½). The eigen-vector, $e(\text{H}) = (0.3550,\ 0.3955,\ 0.7603,\ -0.3736)$, required to solve Eq. (1) related to H-point ordering were extracted from SRO calculation at 1.15T$_{sp}$. The eigen-vectors were taken at 1.15 times above spinodal temperature (T$_{sp}$), where SRO of one of the dominant pairs diverges or `inverse-SRO' goes to zero, i.e., $\boldsymbol{\alpha}_{\mu\nu}^{-1}(\mathbf{k}_o;\text{T}_{sp})$=0.

The estimated occupation at r$_1$=(000) and r$_2$= (½ ½ ½) lattice positions in B2 phase using Eq. (2) are (Cr=0.30, Ni=0.20, Al=0.45, Fe=0.05) and (Cr=0.10, Al=0.05,

Ti=0.20, Fe=0.65) at.-frac., respectively. Similarly, the occupation probability in L2$_1$ phase can be estimated at three symmetry inequivalent sites, namely, r$_1$=(000), r$_2$=(½ ½ ½), and r$_3$=r$_4$=(¼ ¼ ¼) & (¾ ¾ ¾) as (Cr=0.25254, Ni=0.17051, Al=0.31826, Ti=0.20, Fe=0.05869), (Cr=0.14746, Ni=0.02949, Al=0.18174, Fe=0.64131), and (Cr=0.200, Al=0.10.0, Ni=0.250, Ti=0.100, Fe=0.350) at.-frac., respectively.

The concentration-wave analysis combined with direct DFT calculations reflects the energy stability of B2 and L2$_1$ phases in BCC Cr$_{0.20}$Ni$_{0.10}$Al$_{0.25}$Ti$_{0.10}$Fe$_{0.35}$. The formation energy difference of B2 and L21 phases with respect to BCC phase is $\Delta E_{form}(B2 - BCC) = -35.22$ meV-atom$^{-1}$ and $\Delta E_{form}(L2_1 - BCC) = -53.72$ meV-atom$^{-1}$, respectively, which shows that B2 and L2$_1$ phases are energetically more stable than BCC. Notably, L2$_1$ phase is the energetically most stable phase of all three. The SRO calculations in **Fig. 3a** predicted the possibility of L2$_1$ phase, which was recently observed by Wolf-Goodrich *et al.* [**29**]. The H= [111] point instability in Cr$_{0.20}$Ni$_{0.10}$Al$_{0.25}$Ti$_{0.10}$Fe$_{0.35}$ shows B2-type ordering in **Fig. 5a**, which is dominated by Ni-Al SRO pair. The ordering behavior arises from the filling of bonding states that results into strong hybridization as shown in **Fig. 4a** through diffused BSF near Fermi-level. The stronger hybridization in BSF indicates increased charge-fluctuations among various alloying elements in Cr$_{0.20}$Ni$_{0.10}$Al$_{0.25}$Ti$_{0.10}$Fe$_{0.35}$ HEFS. The increased charge-fluctuations at lower temperatures can lead to the development of short-range order, e.g., B2 and L2$_1$ in BCC alloys, which enhances the hybridization among various alloying elements of complex alloy systems.

On the other hand, the formation energy difference of B2 and L2$_1$ ordering phases with respect to disorder (BCC) phase of Cr$_{0.20}$Ni$_{0.10}$Al$_{0.30}$Fe$_{0.40}$ HEFS, i.e., $\Delta E_{form}(B2 - BCC) = -23.53$ meV-atom$^{-1}$ and $\Delta E_{form}(L2_1 - BCC) = -23.55$ meV-atom$^{-1}$, shows that ordering phases are energetically degenerate. This further establishes that adding Ti to Ni-Cr-Al-Fe HEFSs plays a crucial role in stabilizing L2$_1$ phase.

The total density of states (TDOS) is also a good indicator of alloy stability, for example, peak or high density of electronic-states at Fermi-level leads to instability in alloy whereas valley (pseudo-gap) or very-low densities suggest stability [**36**]. We performed electronic-structure calculations of Cr$_{0.20}$Ni$_{0.10}$Al$_{0.25}$Ti$_{0.10}$Fe$_{0.35}$ and Cr$_{0.20}$Ni$_{0.10}$Al$_{0.30}$Fe$_{0.40}$ HEFSs and show the TDOS of disorder (BCC) and partially ordered (B2/L2$_1$) phases in **Fig. 6a-b**. The TDOS of Cr$_{0.20}$Ni$_{0.10}$Al$_{0.25}$Ti$_{0.10}$Fe$_{0.35}$ in **Fig. 6a** shows that both majority-spin and minority-spin channel have a pseudo-gap at the Fermi energy, indicative of increased energy stability [**36**]. The predicted E$_{form}$ of -12.27 meV-atom$^{-1}$ (BCC), -47.49 meV-atom$^{-1}$ (B2), and -65.99 meV-atom$^{-1}$ (L2$_1$) in disorder and partially ordered phases also confirms our analysis. In **Fig. 6b,** the BCC phase shows pseudo-gap region both in up-spin and down-spin channel, which suggests strong stability. The low E$_{form}$ of -21.42 meV-atom$^{-1}$ also confirms our hypothesis. The TDOS of B2 and L2$_1$ phases are identical and show strong pseudo-gap region in up-spin channel, however, down-spin channel shows a peak structure just below the Fermi-level. Unlike TDOS of Cr$_{0.20}$Ni$_{0.10}$Al$_{0.25}$Ti$_{0.10}$Fe$_{0.35}$ HEFS in **Fig. 6a,** the presence of large electronic density of states in Cr$_{0.20}$Ni$_{0.10}$Al$_{0.30}$Fe$_{0.40}$ HEFS in **Fig. 6b** at Fermi-level leads to weaker change in energy stability in B2 (-23.53 meV-atom$^{-1}$) and L2$_1$ (-23.55 meV-atom$^{-1}$) phases compared to disorder phase.

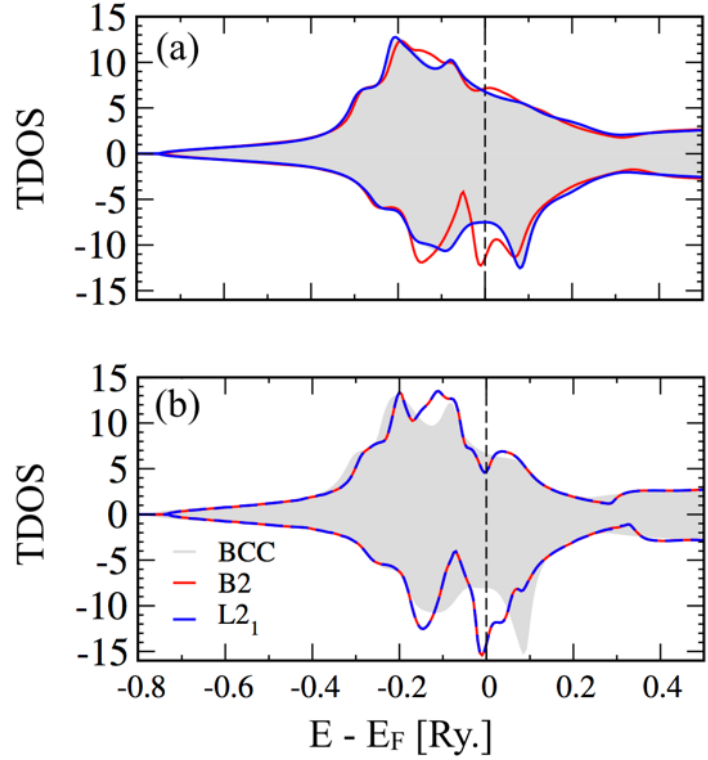

**Figure 6**. Total density of states of (a) $Cr_{0.20}Ni_{0.10}Al_{0.25}Ti_{0.10}Fe_{0.35}$, and (b) $Cr_{0.20}Ni_{0.10}Al_{0.30}Fe_{0.40}$ HEFSs in BCC (grey region), B2 (blue lines), and $L2_1$ (red lines) phases.

**Conclusion**

In summary, the density-functional theory based linear-response theory was used to directly calculate the short-range-order for all atomic pairs simultaneously relative to the homogeneously disordered BCC phase. We showed that the order-disorder transformation, i.e., BCC-to-B2 and BCC-B2-$L2_1$, can be controlled by compositional tuning. The proposed hypothesis of SRO-controlled ordering transformation was exemplified in Cr-Ni-Al-Ti-Fe based ferritic-steels, and we show that the predicted ordering pathways are in good agreement with existing experiments. Our calculations also indicate the possibility of coexistence of ordering phases such as B2 and $L2_1$ below phase decomposition temperature. This study further emphasizes that SRO is important both from fundamental and application point of view as it is known to affect phase selection [16] and mechanical response [18]. Therefore, the tunability of SRO in multi-principal element alloys using purely chemistry provides unique insights for controlling phase transformation, which shows the usefulness of our theory guided design of next generation high-entropy ferritic steels.

**Acknowledgements**

PS would like to thank Dr. Michael Gao (NETL) and JMR for the invitation to contribute to the **JMR Early Career Scholars in Materials Science 2022**. We thank Dr. Marshal Amalraj at Aachen University for fruitful discussions. Work at Ames Laboratory was supported by the U.S. Department of Energy (DOE) Office of Science, Basic Energy Sciences, Materials Science & Engineering Division. Research was performed at Iowa State University and Ames Laboratory, which is operated by ISU for the U.S. DOE under contract DE-AC02-07CH11358.